\documentstyle[12pt]{article}
\begin{document}
\vspace{2cm}
\begin{center}
 \huge  INFLATING DEFECTS
\end{center}   
\vspace{3ex}    
\centerline{\large \em Patricio S. Letelier\footnote{Permanent Address: 
Departamento de Matem\'atica Aplicada--IMECC,
Universidade Estadual de Campinas,
13081 Campinas. S.P., Brazil\\ email: letelier@ime.unicamp.br}
 and Lu\'\i s E. Mendes\footnote{ email: m1480@beta.ist.utl.pt}  }  
\vspace{2ex} 
\begin{center} 
Departamento de F\'\i sica \\ Instituto Superior T\'ecnico\\ Av. 
Rovisco Pais 1\\ 1096 Lisboa, Portugal 
\end{center}

\vspace*{2cm}
 
 \centerline{\small  Abstract}
  \vspace{2ex}
 \baselineskip 0.5cm   

A Kantowski-Sachs 
cosmological model with an O(3) global  defect  as a source is studied
  in the context of  the   topological  inflation scenario. 


\vspace{2ex}
\centerline{ \small July 14 , 1994}  
  \newpage
 \baselineskip 0.7cm

The Universe in which we live is, to a good 
approximation, flat and  highly isotropic and homogeneous on the largest
scales and it is
well  described by the standard 
Friedmann-Lemaitre-Robertson-Walker (FLRW) 
cosmological model .  The realization of  this model
 is only possible  with a very unnatural 
fine-tuning of the initial conditions thus making our universe  
improbable. 
The introduction of a phase of exponential expansion -- inflation -- short 
after the Big-Bang can solve these fine-tuning problems as well as  other
 long standing  puzzling features of our Universe \cite{guth}\cite{kolb}. 

Another problem of a different nature is  the monopole problem that
 arises when we take into account the predictions of Grand Unified 
Theories (GUT's).  Massive monopoles appear   in such large numbers 
during the GUT phase transition  that their  sole contribution  to 
the density of 
the universe would be enough to have caused the recolapse of the universe 
long ago. As such recolapse has not occurred and no monopole has been 
observed yet, we must find  a mechanism to eliminate these
 monopoles. Once again  inflation is the key: the exponential expansion of 
the universe will dilute the density of monopoles thereof solving 
this problem. In fact, it was this so called primordial monopole problem, 
one of the main motivations to introduce inflation \cite{guth}.     

Although inflation can solve the problems of the FLRW model, 
inflationary models also need some amount of fine tuning 
in the initial conditions.  For instance, in the ``new inflationary"
            models, the initial value of the scalar field 
must be localized in a narrow band of width $0.11 m_{pl}$ around 
the top of the potential \cite{piran}.

Recently,   there have been some claims that inflation could occur 
without   fine-tuning in 
the cores of topological defects where the scalar field $\phi$ is in the 
symmetric phase and $\phi=0$ \cite{vilenkin}\cite{linde}. If the defects are 
stable then this  topological inflation will never end. All the models of
 this type have been considered in the context 
of the FLRW models.  Another possibility is to consider unstable defects 
like sphalerons \cite{man} or unwinding textures \cite{hac}.

In this note we extend these ideas by 
constructing an explicit  model where inflation is
 obtained in a Kantowski-Sachs 
model of universe by means of a defect  that in the unbroken phase has
an  O(3)  global symmetry  and whose vacuum manifold is  O(2).  We
 show  (as in \cite{hac})  that even if the defects  are not stable,  an inflationary phase of the universe is also possible.   
Since the  Kantowski-Sachs cosmological
model is  an anisotropic model we   study also its isotropisation along
the inflationary phase.

The  Kantowski-Sachs metric is (c=1),
\begin{equation}
ds^2 = dt^2 - a^2(t) dr^2 - b^2(t) (d \theta^2 + \sin^2 \theta d \varphi^2).
\label{kas}
\end{equation}
The  global O(3)  defect that we shall consider  is the one 
described by the Lagrangian 
density built  with the  triplet of  scalar fields $\Phi^a,$ $(a=1,2,3),$
\begin{equation}
L=\sqrt{-g}[\frac{1}{2}g^{\mu \nu}\partial_{\mu }\Phi^{a} 
\partial_{\nu }\Phi^{b}\delta_{a b} -V(\Phi)] 
\label{lag}
\end{equation}
where $V(\Phi)$ is the usual Mexican hat potential,
\begin{equation}
V(\Phi)\equiv \frac{1}{4}\lambda(\delta_{a b}\Phi^a \Phi^b-\eta^2)^2.
\label{pot}
\end{equation}
 The   O(3) field $\Phi$ has the form 
\begin{equation}
\Phi\equiv(\Phi^a)=\eta\phi(t)[\sin\vartheta \cos\varphi,
\sin\vartheta \sin\varphi, \cos\vartheta].
\label{Phi}
\end{equation}
   The Einstein equations (in units such that  $8\pi G=1$), are
\begin{eqnarray}
2 \frac{\dot{a}}{a} \frac{\dot{b}}{b} + (\frac{\dot{b}}{b})^2 +
\frac{1}{b^2} &=&  
\frac{\eta^2}{2} (\dot{\phi}^2 + 2 \frac{\phi^2}{b^2} + 
\frac{\lambda\eta^2}{2} (\phi^2 - 1)^2)  \nonumber\\ 
2 \frac{\ddot{b}}{b} + (\frac{\dot{b}}{b})^2 + \frac{1}{b^2}& =&
\frac{\eta^2}{2} (- \dot{\phi}^2 + 2 \frac{\phi^2}{b^2} + 
\frac{\lambda \eta^2}{2} (\phi^2 - 1)^2) \nonumber \\ 
\frac{\ddot{a}}{a} + \frac{\ddot{b}}{b} + 
\frac{\dot{a}}{a} \frac{\dot{b}}{b} &=& 
\frac{\eta^2}{2} (- \dot{\phi}^2 +
\frac{\lambda \eta^2}{2} (\phi^2 - 1)^2).
\label{efe}
\end{eqnarray}
The equations for the scalar fields reduce to
\begin{equation}
\ddot{\phi} + (\frac{\dot{a}}{a} + 2 \frac{\dot{b}}{b}) \dot{\phi} + 
\frac{2}{b^2} \phi + \lambda \eta^2 \phi (\phi^2 - 1) = 0
\label{eef}
\end{equation}

Note that this equations are the same as the ones of  a single 
scalar field  \cite{lm}
except for the term $\frac{2}{b^2}\phi$  
of (\ref{eef})  and  $ \eta^2\phi^2 /b^2$ 
on the right hand side of the first 
two components of the Einstein equations (\ref{efe}).
  This last term  alone enters equally in the
 components  $T^{t}_{t}$ and $T^{r}_{r}$
and represents a cloud of Nambu strings with spherical
 symmetry \cite{let79}\cite{barriola}. The Einstein equations 
for the Kantowski-Sachs models for a
cloud of spherically symmetric strings admit a first integral \cite{let83}.

 If we assume that the defect is stable like a  O(3) monopole 
in our case, then forever $\phi = 0$ in the   
core of the defect, the extra term is $0$ and the potential acts 
as a cosmological constant with $p = - \rho$ driving the exponential 
expansion inside the defect. This is the analogous situation to the 
one analyzed by Vilenkin in the context of the FLRW model. Note that in 
this case there is no need for any type of fine tuning, because $\phi=0$ in 
the core of the defects. However, when we take into account the 
fact that global defects are not necessarily stable, e.g., a global texture has dynamical instability,  we can use   
the  defect field itself  to produce slow-roll inflation. 
If we assume that the 
defect vanishes very slowly inside the core, $\dot{\phi} \approx 0$,  
 the field will roll-down the potential also very slowly. During this 
slow-roll, the interior of the defect expands quasi-exponentially. 
During this process the extra term from the defect has a 
negligible role. In fact, in the beginning of inflation $\phi=0$ and 
this term vanishes; short after inflation has begun $b$ starts an  
exponential growth and the extra term rapidly becomes negligible. 
The realization of this idea is depicted in Fig.1 in which
 we compare this inflating 
defect  model with usual inflation obtained with the same potential (in 
the simulations we used the variables $\alpha=\ln a$ and 
$\beta = \ln b$). We can see that the presence of this extra term can 
increase slightly the 
expansion during the inflationary epoch. This can be understood from the fact 
that the  term    $ \eta^2\phi^2 /b^2$ is always positive and  
whenever it appears  it adds to the potential. 

As in the models with this type of potentials, after reaching 
the bottom of the potential $\phi$ oscillates around the minimum 
of the potential giving rise to the reheating of the universe. In this
model the slow-roll condition plays an important role,
we need that the defect decays with a suficiently large 
time scale to have enough time to produce inflation in 
the very early universe. 
In other words we need some fine 
tunning of the initial conditions.
Maintaining all the initial conditions fixed, we can increase the time 
scale for the decay increasing the value of $\eta$. This is because 
the larger the value of $\eta$, the far from the origin the minimum 
of the potential will be.

A good measure of the anisotropy of the model is given by the ratio 
\begin{equation}
\frac{ \sigma}{\theta}=\frac{\frac{\dot{a}}{a}
 -\frac{\dot{b}}{b}}{ \frac{\dot{a}}{a} +2\frac{\dot{b}}{b}}
\label{shear}
\end{equation}
where $\sigma$ is the shear scalar and $\theta$ the expansion rate. 
We find that in our model this quantity goes to zero very fast, 
as in the model with a single scalar field \cite{lm}, thus indicating 
that once inflation has started the model rapidly isotropizes. \\

 \noindent {\bf Acknowledgments} \\

 PSL thanks the hospitality of Departamento de F\'\i sica do Instituto 
 Superior
 T\'ecnico. LEM was supported by Junta Nacional de 
 Investiga\c c\~{a}o Cient\'\i fica e 
 Tecnol\'ogica (JNICT). PSL also thanks  discussions with P. Shellard and 
A. Vilenkin.

\newpage
FIGURE CAPTIONS\\

FIG. 1 The solid curve shows 
the evolution of $R=(a b^2)^{1/3}$ in our model 
during the last stages of inflation while the dashed curve represents 
the same quantity for the model with a single scalar field obtained 
with the same initial conditions. We see that the presence of the extra term 
in the model with the monopole can increase slightly the expansion during 
the inflationary phase.The initial conditions are $\alpha=1$, 
$\dot{\alpha}=2.25 \times 10^{11}$, 
$\beta=0.17$, $\dot{\beta}=5 \times 10^{-14}$, $\phi=0$, and 
$\dot{\phi}=10^{-43}$; we have set $\lambda=1$ and $\eta=1.3$. \\

FIG. 2 After reaching the bottom of the potential, the scalar 
field oscillates around the minimum. The initial conditions are the same as 
in Fig. 1 

\end{document}